\documentclass[12pt]{JHEP3} 

\usepackage{epsfig,multicol,bbm}
\usepackage{latexsym}
\usepackage{amssymb}
\usepackage{amsfonts}
\usepackage{amsmath}
\usepackage{bm}
\usepackage{times}
\usepackage{units}

\newcommand{\cf}{\textit{cf.}~}
\newcommand{\ie}{\textit{i.e.,}~}

\newcommand{\sgr}{{\rm Sgr~A}^*}
\newcommand{\Msgr}{M_{\text{Sgr A}^*}}

\title{Constraining scalar fields with stellar kinematics and collisional dark matter}

\author{Pau Amaro-Seoane
                \\
        Max-Planck Institut f{\"u}r Gravitationsphysik,
        Albert-Einstein-Institut, Potsdam, Germany and\\
        Institut de Ci{\`e}ncies de l'Espai (CSIC-IEEC), Bellaterra, Barcelona, Spain\\
        }

\author{Juan Barranco \& Argelia Bernal\\
        Max-Planck Institut f{\"u}r Gravitationsphysik,
        Albert-Einstein-Institut, Potsdam, Germany\\
        }

\author{Luciano Rezzolla\\
        Max-Planck Institut f{\"u}r Gravitationsphysik,
        Albert-Einstein-Institut, Potsdam, Germany and\\
        Department of Physics and Astronomy
        Louisiana State University, Baton Rouge, LA, USA
        }

\received{\today}               
\revised{}
\accepted{}             

\abstract{
The existence and detection of scalar fields could provide solutions
to long-standing puzzles about the nature of dark matter, the dark
compact objects at the centre of most galaxies, and other phenomena.
Yet, self-interacting scalar fields are very poorly constrained by
astronomical observations, leading to great uncertainties in estimates
of the mass $m_\phi$ and the self-interacting coupling constant
$\lambda$ of these fields. To counter this, we have systematically
employed available astronomical observations to develop new
constraints, considerably restricting this parameter space. In
particular, by exploiting precise observations of stellar dynamics at
the centre of our Galaxy and assuming that these dynamics can be
explained by a single boson star, we determine an upper limit for the
boson star compactness and impose significant limits on the values of
the properties of possible scalar fields. Requiring the scalar field
particle to follow a collisional dark matter model further narrows
these constraints. Most importantly, we find that if a scalar dark
matter particle does exist, then it cannot account for both the
dark-matter halos and the existence of dark compact objects in
galactic nuclei
}


\begin{document}


\section{Introduction}
\label{introduction}

A large number of astronomical observations little doubt about the
existence of dark matter (DM) and its dominant role in the matter
composition in the Universe~\cite{Bertone:2004pz}. An even larger
number of candidates has been proposed to play the role of this
unknown component.  Among those candidates, ultra-light scalar fields
have been extensively studied, providing a robust paradigm for DM
particles~\cite{Hu:2000ke,sflocal2000,sflocal2001,sflocal2004} and
dark energy~\cite{Frieman:1995pm}. Another scalar field, the inflaton,
is a crucial ingredient in inflationary cosmology~\cite{Guth:1980zm,
  Linde:1981mu, Albrecht:1982mp}. The success of cosmological and
astrophysical models that invoke scalar fields has, in addition,
motivated further studies of self-gravitating systems made of scalar
fields, such as non-topological solitons~\cite{Lee:1986ts},
oscillatons~\cite{Seidel:1991zh,UrenaLopez:2001tw,Alcubierre:2003sx}
and boson stars (BS)~\cite{Jetzer92,Schunck:2003kk}.

In particular, when considering a BS the scalar field is associated
with a spin-zero boson with mass $m_\phi$ and self-interacting
coupling constant $\lambda$.  The essentially unconstrained freedom in
choosing the free parameters characterizing the scalar field (namely
its mass and self-coupling constant), allows one to build BS-models
that can account for many different astrophysical objects: from
galactic dark matter halos~\cite{arbey1, arbey2, Matos:2007zza,
  Arbey:2003sj}, to dark compact objects (DCOs) like
MACHOS~\cite{HernandezEtAl04,Barranco:2010ib} and black hole
candidates~\cite{Torres:2000dw, Guzman:2005bs,Guzman:2009zz}.
Clearly, the only way to restrict the very large space of parameters
spanned by BS models is by exploiting astronomical observations and
use them to set constraints on the properties of the scalar
field. This is, in essence, the goal of this paper.

The recent advances in high-angular resolution instrumentation have
provided the possibility to study the central regions of galaxies with
unprecedented precision. Space-borne telescopes, such as the Hubble
Space Telescope, and ground observations using adaptive optics have
opened a new window on the innermost central regions of galactic
nuclei. Particularly striking is the present ability to study the
kinematics of stars or gas in regions of milliparsec scale for the
Milky Way~\cite{GhezEtAl03,SchoedelEtAl03} and of sub-parsec scale for
external galaxies~\cite{FF04,MoranEtAl99, Kormendy03}. One of the most
intriguing conclusions of these observations is that dark compact
objects (DCOs), with masses ranging between $M_{\rm DCO}\simeq
10^6-10^9\,M_{\odot}$, are hosted at the centre of most non-active
galaxies.

In the case of our own galactic centre, the study of the innermost
stellar dynamics has provided very convincing evidence for the
existence of a DCO associated with the radio point source Sagittarius
A$^*$ ($\sgr$). The closest Keplerian orbits examined are those of the
so-called ``S-stars'' (also referred to as ``SO-stars''). By following
the orbit of one of them, \ie the star S2 (also referred to as SO2),
the mass of $\sgr$ has been estimated to be about $3.7\times
10^6\,M_{\odot}$ within a volume with radius no larger than $6.25$
light-hours~\cite{SchoedelEtAl03,GhezEtAl03b}. More recent data based
on $16$ years of observations has reduced the uncertainty and set the
mass of the central DCO to $\sim 4.1 \pm 0.6 \times 10^6 \,
M_{\odot}$~\cite{EisenhauerEtAl05, GhezEtAl08, GillessenEtAl09}.

Assuming that the DCO in $\sgr$ is a BS, we have used the
observational restrictions on the volume and mass to set substantial
constraints on the mass of the scalar field and on its
self-interacting coupling constant. The approach followed using the
observational data for $\sgr$ can be applied to any other galaxy for
which there is an accurate measurement of the mass and dimensions of
the central object. As a result, we have also considered the case of a
nearby galaxy, NGC 4258, obtaining additional constraints, which set
much tighter limits when considered in combination with those of
$\sgr$.

Finally, we have considered several DM candidates and discussed how
present astronomical observations can be exploited to set additional
limits on the scalar-field properties invoked by these DM
models~\cite{Hu:2000ke, arbey1, arbey2, Matos:2007zza,
  Arbey:2003sj}. While some of these constraints overlap with those
obtained when considering BSs, others cover a distinct region of the
space of parameters and thus imply that a scalar field that could
explain the rotation curves in nearby galaxies cannot be the same
composing a BS representing the DCOs in galactic nuclei.

The structure of the paper is the following one. In
Sect.~\ref{BSgeneral} we briefly review the properties of BSs in the
two regimes of weak and strong self-interaction. Section~\ref{mcfabs}
is instead dedicated to the issue of the maximum compactness of a BS,
where we show that the definition of an effective maximum compactness
for a BS is not only possible, but actually useful when associating
BSs to DCOs in galactic centres. In Sect.~\ref{results} we therefore
derive our constraints on the scalar-field properties using both
considerations on BSs models and of DM models. Finally, our
conclusions are contained in Sect.~\ref{conclusions}.

\section{Boson star generalities}
\label{BSgeneral}
Boson stars are solutions to the Einstein Klein-Gordon system of equations
\begin{equation}\label{ekg}
G_{\mu\nu}=8\pi G T_{\mu\nu}\,, \qquad 
\left( \Box - \frac{dV}{d\Phi}  \right )=0\,, 
\end{equation}
where $\Box\equiv(1/\sqrt{-g})\partial_{\mu}[\sqrt{-g}g^{\mu
    \nu}\partial_\nu]$. We have adopted natural units $\hbar=c=1$, and
hence $G=1/m_{_{P}}^{2}$, where $m_{_{P}}$ is the Planck mass.  The
stress energy tensor $T_{\mu\nu}$ corresponds to a complex scalar
field minimally coupled to gravity and with a scalar potential
$V(|\Phi|^2)$,
\begin{equation}
\label{tensor}
T_{\mu\nu}=\frac{1}{2}[\partial_\mu \Phi^*\partial_\nu\Phi+
\partial_\mu\Phi\partial_\nu\Phi^*]-
\frac{1}{2}[\Phi^{*,\alpha}\Phi_{,\alpha}+V(|\Phi|^2)]\,.
\end{equation}
We will restrict ourselves to the case where the potential of the
scalar field is given by $V=\frac{1}{2}m_\phi^2|\Phi|^2
+\frac{\lambda}{4}|\Phi|^4$, where $m_\phi$ is the mass of the scalar
field and $\lambda$ its self-interaction coupling constant.
Originally BSs were introduced by Kaup~\cite{Kaup:1968zz} and studied
later by Ruffini and Bonazzola~\cite{Ruffini:1969qy}, who considered
just the mass term in the scalar potential. The case of
self-interacting scalar fields were later discussed by Colpi and
collaborators in~\cite{Colpi:1986ye}. If for simplicity the scalar
field is assumed to be spherically symmetric and with a harmonic time
dependence, $\Phi(r,t)=\phi(r) e^{-i \omega t}$, the space-time part
is static. Considering spherical symmetry, the line element can be
written as $ds^2=-B(r)dt^2+A(r)dr^2+r^2d\Omega^2$ and the Einstein
Klein-Gordon equations reduce to a system of ordinary differential
equations for the metric functions $A$ and ${\tilde B} \equiv
{m^2_{\phi} B}/{\omega^2}$, and for the scalar field
$\phi$~\cite{Colpi:1986ye}.  Additional quantities appearing in the
equations are $\sigma\equiv\sqrt{4\pi /m_{_{P}}^2}\phi$, which reduces
to the density in the Newtonian limit, and the dimensionless
self-interacting coupling constant~\cite{Colpi:1986ye}
\begin{equation}
\label{Lambda}
\Lambda\equiv \frac{\lambda m_{_{P}}^2}{4 \pi m_\phi^2}\,. 
\end{equation}

The freedom of choice of $\lambda$ and $m_\phi$ is reflected in the
freedom of choosing $\Lambda$ and it has been shown that even when
$\lambda \ll 1$, the resulting configuration may differ significantly
from the non-interacting case~\cite{Colpi:1986ye}. For sufficiently
small $\lambda$, on the other hand, self-interaction may only be
ignored if $\lambda \ll m_\phi^2/m_{_{P}}^2$, that is, if $\Lambda \ll
1$. On the other hand, configurations with $\Lambda \gg 1$ are
interesting as their masses may be comparable to those of their
fermion counterparts if $\lambda \sim 1$. This illustrates the
relevance of taking both regimes, $\Lambda \sim 1$ and $\Lambda \gg 1$
in the analysis of BS configurations. In the next two subsections we
review them briefly for completeness and to introduce a notation that
will be useful later on.

\FIGURE[t]{
\includegraphics[angle=0,width=0.475\linewidth]{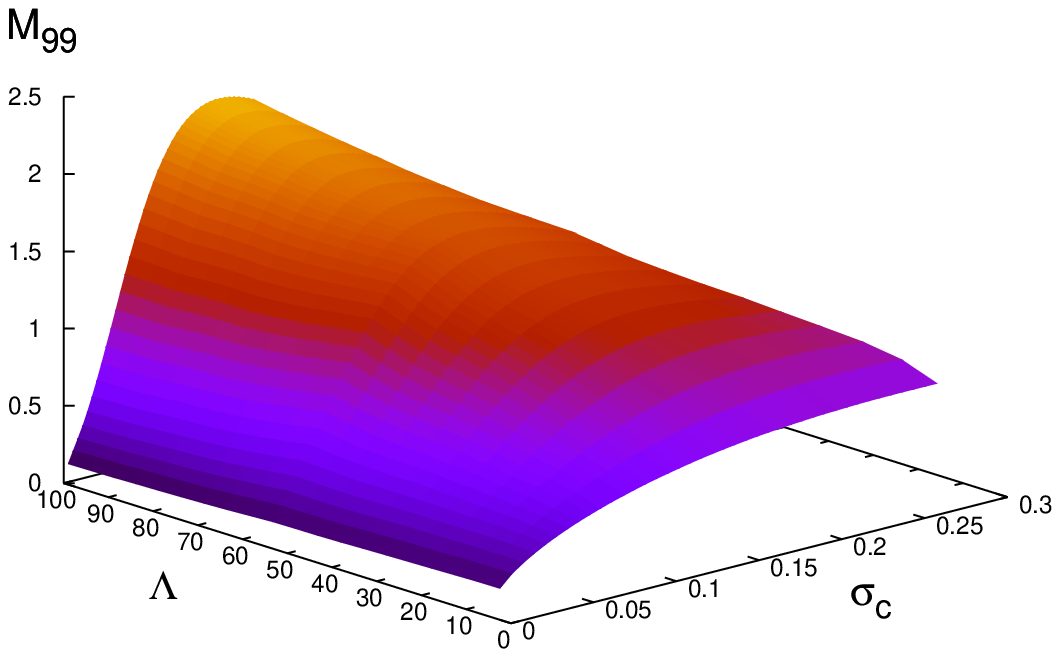}
\hskip 0.5cm
\includegraphics[angle=0,width=0.475\linewidth]{./figs/masaBS}
\\
\caption{
  Gravitational mass $M_{99}$ of equilibrium configurations of
  BSs for different values of $\sigma_{\rm c}$ and $\Lambda$ in the
  regime $\Lambda_3 \ll 1$; black squares indicate configurations with
  the maximum mass. The left panel, in particular, shows a
  three-dimensional view of the space of solutions, while the right
  panel offers a view of three slices given values of $\Lambda$.
\label{equilibrio}
}}

\subsection{Weak self-interaction ($\Lambda_{3} \ll 1$)}
\label{BSsmalllambda}

We next we derive equilibrium configurations for the Einstein
Klein-Gordon system in the weak self-interaction regime, \ie when
\mbox{$\Lambda_{3} \equiv \Lambda/1000 \ll 1$}, following a procedure
similar to the one discussed before in~\cite{Jetzer92,
  Schunck:2003kk, Guzman:2005bs, Guzman:2009zz}. In particular, in
this case the Einstein Klein-Gordon system of equations can be solved
as an eigenvalue problem for the radial metric function at the BS'
centre ${\tilde B}_{\rm c}$ after specifying a finite but arbitrary
value for the ``central density'' $\sigma_{\rm c}$. Additional
boundary conditions are those of regularity at the centre, \ie $A_{\rm
  c}=1$, $\sigma'_{\rm c}=0$, where the prime denotes a derivative
with respect to the new dimensionless radial coordinate $x \equiv r
m_\phi$, and of asymptotic flatness, \ie $\sigma(x=\infty)=0$.

The system is solved numerically for different values of $\sigma_{\rm
  c}$ and $\Lambda$ using a standard shooting method (given a
$\sigma_{\rm c}$, there is a unique value of ${\tilde B}_{\rm c}$ for
which the boundary conditions are satisfied). Because we use a finite
numerical domain, the condition for $\sigma$ at infinity is in fact
demanded at the outermost point of the domain $x_{\rm out}$, and the
shooting procedure is performed for different values of $x_{\rm
  out}$. As $x_{\rm out}$ increases, the shooting parameter ${\tilde
  B}_{\rm c}$ converges, and we choose the solution by requiring that
the condition $\sigma(x_{\rm out})=0 $ holds within a prescribed
tolerance (\ie $\sigma(x_{\rm out}) \lesssim 10^{-9}$).

Once the equilibrium solution is obtained, its mass is computed simply
as
\begin{equation}
\label{mass}
M \equiv \frac{x_{\rm out}}{2}\left(1-\frac{1}{A(x_{\rm out})}\right)\,.
\end{equation}
We note that since the scalar field $\sigma$ decays exponentially, the
radius of the star is, at least formally, not finite. In practice,
however, we define $M_{99} \equiv 0.99\,M$, that is as $99\%$ of the
total gravitational mass $M$; as a result, we also define $R_{99}$ as
the radius containing $M_{99}$ and set this to be the
\textit{``effective radius''} of the BS. As a final remark we note
that in view of the units adopted, both quantities $M$ and $R$ are
dimensionless variables. The physical units for the mass $\hat{M}$ and
radius $\hat{R}$ can be recovered by using the following relations
\begin{equation}
\label{units}
\hat{M}=M\frac{m_{_{P}}^2}{m_\phi}\,, \qquad \
\hat{R}=R\frac{\hbar}{m_\phi}\,. 
\end{equation}

Figure~\ref{equilibrio} reports the gravitational mass $M_{99}$ of
equilibrium configurations of BSs for different values of $\sigma_{\rm
  c}$ and $\Lambda$ in the regime $\Lambda_3 \ll 1$. In particular,
the left panel shows a three-dimensional view of the space of
solutions, while the right panel offers a view of three slices given
values of $\Lambda$. In both panels, black squares indicate
configurations with the maximum mass, thus distinguishing stable
solutions (to the left of the squares) from unstable ones (see
discussion in Sect.~\ref{sobs}).

\subsection{Strong self-interaction ($\Lambda_{3} \gg 1$)}
\label{BSstrong}

\FIGURE[t]{
\centerline{
\includegraphics[width=8.0cm]{./figs/maxmassnew}}
\caption{Gravitational mass $M_{99}$ of equilibrium configurations of
  BSs for different values of $\sigma_{\rm c}$ in the regime
  $\Lambda_3 \gg 1$. Also in this case the black square indicates the
  configurations with the maximum mass.
\label{lambdabig}}
}

\FIGURE[t]{
\includegraphics[angle=0,width=0.475\linewidth]{./figs/compactBS}
\hskip 0.5cm
\includegraphics[angle=0,width=0.475\linewidth]{./figs/compactnessnew}
\\
\caption{
Effective compactness $C\equiv M_{99}/R_{99}$ for a spherical
  BS shown as a function of the central density $\sigma_{\rm
    c}$. Shown in the left panel are the values in the regime of weak
  self-interaction $\Lambda_3 \ll 1$, while the right panel reports
  the values in the regime of strong self-interaction $\Lambda_3 \gg
  1$. Also in this case the black squares in both panels indicates the
  configurations with the maximum mass.
\label{fig.Compactness}
}}

The solution of the Einstein Klein-Gordon equations is much simpler
when considering instead the strong self-interaction regime, \ie when
\mbox{$\Lambda_{3} \gg 1$}. In this case, in fact, introducing the new
dimensionless variables~\cite{Colpi:1986ye}
\begin{eqnarray}
\label{reparameter}
\sigma_*&=&\Lambda^{1/2}\sigma \,, \\
x_*&=&\Lambda^{-1/2}x\,.
\end{eqnarray}
and neglecting terms of order $O(\Lambda^{-1})$, the Klein-Gordon
equation can be solved algebraically to yield~\cite{Colpi:1986ye}
\begin{equation}\label{sigmanew}
\sigma_*=\left(\frac{1}{{\tilde B}}-1\right)^{1/2}\,.
\end{equation}
so that ${\tilde B}_{\rm c}=1/(\sigma^2_{*,\rm c}+1)$. Furthermore,
the full set of the Einstein Klein-Gordon equations does not depend on
$\Lambda$ at first order and can be solved much more easily. As in the
weak self-interaction regime, equilibrium configurations are obtained
solving a set of ordinary differential equations with the same
boundary conditions. One important difference, however, is that in
this limit the density $\sigma$ does go to zero at a finite radius
$x_{\text{S}}$ and, as for ordinary fluid stars, the gravitational
mass $M$ is defined by Eq.~(\ref{mass}) with $x_{\rm
  out}=x_{\text{S}}$. In analogy with the previous treatment, we
define the effective radius of the BS as $R_{99}$, namely as the one
containing $99\%$ of the total gravitational mass\footnote{Note that
  since in the case $\Lambda_3 \gg 1$ the BS has a finite radius, the
  definition of $R_{99}$ is strictly speaking not necessary, but it is
  nevertheless useful to maintain consistency with the case $\Lambda_3
  \ll 1$.}. Finally, because the introduction of the new variable
$x_*$, see eq (\ref{reparameter}), the physical units can be recovered
by using the following relations
\begin{equation}
\label{unitsbiglambda}
\hat{M}=\Lambda^{1/2}M\frac{m_{_{P}}^2}{m_\phi}\,, \qquad 
\hat{R}=\Lambda^{1/2}R\frac{\hbar}{m_\phi}\,. 
\end{equation}

Shown in Fig.~\ref{lambdabig} is the gravitational mass $M_{99}$ of
equilibrium configurations of BSs for different values of $\sigma_{\rm
  c}$ in the regime $\Lambda_3 \gg 1$. Also in this case the black
square indicates the configurations with the maximum mass, thus
distinguishing stable solutions (to the right of the square) from
unstable ones (see discussion in Sect.~\ref{sobs}).

As a final remark, we note that a general analytic expression for the
mass and compactness of a BS in the strong self-interaction regime can
be given and that fits well the data presented in the right panel of
Fig.~\ref{lambdabig}. Such expressions are useful as they allow one to
obtain simple estimates to be used in astronomical observations.

\subsection{Stability of Boson Stars}
\label{sobs}

Clearly, the stability of BS configurations is a basic requirement when
considering them as suitable models of astrophysical objects.
Numerical~\cite{SeidelSuen90,HawleyChoptuik00,Guzman:2004jw} and
analytical~\cite{GleiserWatkins89,LeePang89,KusmartsevMielkeSchunck91}
stability studies of BSs agree on a general result: nodeless BS configurations
are stable under finite radial perturbations if their scalar field central
value $\sigma_{\rm c}$ is smaller than a critical value $\sigma_{\rm crit}$;
configurations with $\sigma_{\rm c} >\sigma_{\rm crit}$, in fact, are either
collapse to a BH or disperse at infinity \cite{Guzman09}.

Note that, for any given value of $\Lambda$, the critical density
$\sigma_{\rm crit}$ also marks the maximum mass corresponds the
maximum mass of the BS, \ie
\begin{equation}
M_{\rm max}(\Lambda) \equiv M(\sigma_{\rm c}=\sigma_{\rm crit}, \Lambda)\,.
\end{equation}

In the regime of $\Lambda_3 \ll 1$, the maximum mass clearly increases
as $\Lambda$ increases as shown in the two panels of
Fig.~\ref{equilibrio}. On the other hand, when $\Lambda_{3} \gg 1$,
one can exploit the fact that the system of the Einstein Klein-Gordon
equations can be identified with that of a perfect-fluid star for
which the effective equation of state is
\begin{equation}\label{eos}
p=\frac{4}{9}\rho_0 \left[\left(
1+\frac{3\rho}{4\rho_0}\right)^{1/2}-1\right]^2\,,
\end{equation}
\noindent
where
\begin{equation}
\rho=\frac{1}{4}\left(\frac{3}{{\tilde B}}+1\right)
\left(\frac{1}{{\tilde B}}-1\right)\,, 
\end{equation}
and $\rho_0=m_\phi^4/\lambda$. Note that in the limit $\rho \ll
\rho_0$ (which is equivalent to $P \ll \rho$ \ie the Newtonian limit)
Eq.(\ref{eos}) reduces to $P={\rho^2}/({16\rho_0})$, thus representing
the equation of state of an $n=1$ polytrope. As a result, the
well-known theorems of stability of fluid stars may be applied to
determine the stability of the BSs. In particular, BSs with $1/{\tilde
  B}_{\rm c}$ smaller than a critical value $1/{\tilde B}_{\rm crit}$
are stable, while those for which $1/\tilde B_{\rm c} > 1/{\tilde
  B}_{\rm crit}$ will either collapse to a black hole or disperse to
infinity. In practice, when $\Lambda_3 \gg 1$, the maximum mass is
$M_{\max}\simeq 0.22$ and is attained at $\sigma_{\rm crit}\simeq
0.97$; this is shown with a black square in Fig.~\ref{lambdabig}.

\section{Maximum Compactness for a Boson Star}
\label{mcfabs}

Since we are interested in modelling astrophysical DCOs as stable BSs,
it is important to determine whether or not a BS admits a maximum
compactness, just as it is the case for an ordinary relativistic
star. At first it may appear that this question is not even well posed
as BSs are in principle infinitely extended and hence always with a
zero compactness. However, even an infinitely extended BS is in
practice centrally condensed, \ie with a \textit{finite} effective
radius $R_{99}$. Hence, it is mathematically legitimate and
astrophysically reasonable to consider whether the \textit{``effective
  compactness''} of a BS, defined as
\begin{equation}
\label{compactness_99}
C(\sigma_{\rm c}, \Lambda)\equiv\frac{ M_{99}(\sigma_{\rm c},\Lambda)}{R_{99}}\,,
\end{equation}
and thus measuring the compactness of a BS with mass $M_{99}$ in a
radius $R_{99}$, can be nonzero and is even upper bounded. As we will
show in what follows, this is indeed the case and $C(\sigma_{\rm c},
\Lambda) \lesssim 0.16$ for all values of $\Lambda$.

Before entering the discussion of the different regimes of
self-interaction it is interesting to note that if computed in terms
of the physical values for the effective mass ${\hat M}_{99}$ and
effective radius ${\hat R}_{99}$, the compactness $C$ is independent
of $m_\phi$ [(\cf Eqs~(\ref{units}) and (\ref{unitsbiglambda})].

In the case of weak self-interaction $\Lambda_{3} \ll 1$, the
compactness (\ref{compactness_99}) is shown in the left panel of
Fig.~\ref{fig.Compactness} for different values of $\Lambda$. Note
that the compactness is a growing function of $\sigma_{\rm c}$, but
also that it has a local maximum. The latter, however, is reached for
unstable configurations, since it refers to BSs whose central density
$\sigma_{\rm c}>\sigma_{\rm crit}$. Shown in fact with black squares
are the configurations with the maximum mass and these are all to the
left of the local maxima: these models mark therefore the BSs with the
maximum compactness for a given choice of central density and coupling
constant, \ie $C_{\rm max}(\sigma_{\rm c}, \Lambda)$.

A very similar behaviour is shown by the compactness in the right
panel of Fig.~\ref{fig.Compactness}, which however refers to the
regime of strong self-interaction $\Lambda_{3} \gg 1$. The obvious
difference in this case is that the compactness does not depend on
$\Lambda$ (and hence on the free parameters of the scalar field
$m_\phi$ and $\lambda$) and attains a maximum value for stable
configurations $C_{\max, \Lambda_{\infty}} = 0.158496114$ at
$\sigma_{\rm crit}\simeq 0.97$; this is shown with a black square in
the right panel Fig.~\ref{fig.Compactness}~\footnote{Note that
  although $C_{\max, \Lambda_{\infty}}$ does not consider the whole
  mass, the corresponding compactness $0.158496114$ is very close to
  the maximum one obtained when considering the whole mass and the
  whole radius and which is given by $M_{\text{S}}/R_{\text{S}} =
  0.159807753$.}. Interestingly, this maximum compactness is close to
that of a typical neutron star with mass $M_{\text{NS}} =
1.4\,M_{\odot}$ and a radius of $R_{\text{NS}} = 12\,{\rm km}$.

Comparing the two panels of Fig.~\ref{fig.Compactness} it is easy
realize that the maximum compactness increases with $\Lambda$ and has
as asymptotic value for $\Lambda \to \infty$ the one given by
\mbox{$C_{\max, \Lambda_{\infty}}$}. This is clearly shown in
Fig.~\ref{maxcompactness}, where we report the maximum compactness
$C_{\rm max}$ as a function of the of coupling constant
$\Lambda$. Shown with a dashed line is the asymptotic maximum
compactness for a spherical BS \mbox{$C_{\max, \Lambda_{\infty}}\simeq
  0.158$}, while shown with a dotted and dot-dashed lines are the
maximum compactnesses for a spherical star \mbox{$C_{\rm max,
    star}=4/9$}~\cite{Buchdahl:59} and the compactness of a black hole
$C=1/2$.

Overall, Fig.~\ref{maxcompactness} shows two important results. The
first one is that although BSs are in principle infinitely extended
they can be considered to have an effective compactness which is upper
limited. The second one is that such a maximum compactness is smaller
than the corresponding one for a fluid star and, of course, of a black
hole. To the be best of our knowledge neither of these two results was
discussed before in the literature.

\FIGURE[t]{ 
\centerline{
\includegraphics[width=8.0cm]{./figs/compactness-lambda}}
\caption{Maximum compactness $C_{\rm max}$ shown as a function of the
  of coupling constant $\Lambda$. Indicated with dashed line is the
  asymptotic maximum compactness for a spherical BS \mbox{$C_{\max,
      \Lambda_{\infty}}\simeq 0.318$}, while a dotted and dot-dashed
  lines are the maximum compactnesses for a spherical start
  \mbox{$C_{\rm max, star}=4/9$} and the compactness of a black hole
  \mbox{$C_{\text{BH}}=1/2$}, respectively.}
\label{maxcompactness}
}

\section{Constraining $m_\phi$ and $\lambda$}
\label{results}

In what follows we discuss how to make use of the astronomical
observations of galactic centres ($\sgr$ and NGC 4258) and of present
DM models to constrain the space of possible parameters for the scalar
field mass $m_{\phi}$ and self-interaction coupling constant
$\lambda$. It is however important to recall that present constraints
on $m_{\phi}$ and $\lambda$ are extremely loose. Previous works that
considered BSs as DCOs candidates~\cite{Guzman:2005bs,Guzman:2009zz},
in fact, neither $m_\phi$ or $\lambda$ were related to an existing
particle or particle candidate. In Ref.~\cite{Torres:2000dw}, on the
other hand, some first estimates were presented but not using the
astronomical observations considered here.

\subsection{Constraints from $\sgr$}

\FIGURE[t]{
\includegraphics[angle=0,width=0.475\linewidth]{./figs/fig5a}
\hskip 0.5cm
\includegraphics[angle=0,width=0.475\linewidth]{./figs/fig5second}
\\
\caption{
  Constraints on the possible values of the mass and coupling
  constant of the complex scalar field. The left panel refers to the
  weak self-interaction regime of $\Lambda_3 \ll 1$, where the dashed
  line refers to the condition (\ref{mphi_max}), while the solid one
  to the condition (\ref{mphi_min}). The shaded region is clearly the
  union of the inequalities. The right panel is the same as the left
  one, but refers to the strong self-interaction regime of $\Lambda_3
  \ll 1$, so that where the dashed line refers to
  (\ref{large_lambda_SgrA_max}), while the solid one to
  (\ref{large_lambda_SgrA_min}). In both panels shown with the insets
  are the same data but in terms of the dimensionless self-interacting
  coupling constant $\Lambda$.
\label{constraints}
}}

A first set of constraints can be set by exploiting the observations
about the dynamics of the S-stars around $\sgr$ and in particular of
the smallest measured periastron for this group of stars. This is
attained for the star S2, whose periastron is $R_{\rm S2} \simeq 17$
light-hour $\simeq 5.9\times 10^{-4}$ pc. If a DCO is indeed hosted at
the centre of $\sgr$, then its radius cannot be larger than $R_{\rm
  S2}$ or it would perturb considerably these orbits. Hence its
minimum compactness is
\begin{equation}
\label{Cmin}
C_{\rm min}=\frac{\Msgr}{R_{\rm S2}} \simeq \frac{1}{3015}\,, 
\end{equation}
where the second equality has been obtained using the measured mass of
$\sgr$, which has been estimated to be \mbox{$M_{\text{Sgr A}^*}=4.1
  \pm 0.6 \times 10^6\,M_\odot$}~\cite{GenzelEtAl10}\footnote{We note
  that larger compactnesses (indeed as large as the ones corresponding
  to a black hole) have been suggested for $C_{\rm min}$ on the basis
  of radio observations at wavelengths of $3.5$ mm and $7$
  mm~\cite{DoelemanEtAl08, DoelemanEtAl09}. The measurements are
  particularly complex and may be contaminated by instrumental errors
  which are difficult to remove completely. In view of these
  uncertainties we prefer to use the more conservative but also more
  accurate estimate (\ref{Cmin}).}. Note that when expressed in terms
of the Schwarzschild radius corresponding to $\sgr$, \ie $R_{\text{S}}
= 2\Msgr$, the periastron of the S2 star is $R_{\text{S2}} \simeq
1520\, R_{\text{S}}$.

The observational evidence that $\sgr$ is a black hole is indeed very
convincing, but we here assume, for the sake of argument, that it is
rather a BS of mass $\Msgr$. In this case, its compactness could be in
the range
\begin{equation}
\label{Crange}
3.32\times 10^{-4} \simeq C_{\rm min} 
\leq C_{\text{BS}}  \leq 
C_{\rm max} \simeq 0.158\,.
\end{equation}
In other words, by changing $m_{\phi}$ and $\lambda$ it is possible to
construct infinite BS models that would be compatible with the
observations of the S-stars and with radii between $R_{\text{S2}}$ and
$R_{\rm min} \equiv \Msgr /C_{\rm max}$. In this way, the condition
(\ref{Crange}) can be used effectively to constrain the space of
parameters of potential scalar fields.

The procedure adopted in practice to enforce the condition
(\ref{Crange}) is somewhat involved and, as in the previous Sections,
we will consider separately the cases of weak and strong
self-interaction. In particular, when $\Lambda_{3}\ll 1$ the
\textit{maximum} value of $m_{\phi}$ compatible with the mass of
$\sgr$ can be set by recalling that the relation between physical and
non-physical masses (\ref{units}) states that the scalar-field mass is
proportional to the BS mass for any given value of the physical
mass. Hence, after fixing $\Msgr$, the mass of the scalar field will
have to satisfy the inequality
\begin{equation}
\label{mphi_max} 
m_\phi \leq \frac{M_{\rm max}(\Lambda)}{\Msgr}\, m_{_{P}}^2\,.
\end{equation}
The condition (\ref{mphi_max}) when the equality holds is shown as a
dashed line in the left panel of Fig.~\ref{constraints}.

Similarly, a lower limit on $m_{\phi}$ can be found by determining,
for each $\Lambda$, the central density $\sigma_{\rm c}^*$ and the
radius $R(\sigma_{\rm c}^*)$ of the BS model having the minimum
compactness, $C_{\rm min}$ (we recall that the compactness is a
monotonic function of $C$). Defining then the minimum mass as $M_{\rm
  min}(\Lambda)\equiv C_{\rm min} R(\sigma_{\rm c}^*)$, we can
constrain the mass of the scalar field to satisfy the inequality
\begin{equation}
\label{mphi_min} 
m_\phi \geq \frac{M_{\rm min}(\Lambda)}{\Msgr}\, m_{_{P}}^2\,.
\end{equation}
The condition (\ref{mphi_min}) when the equality holds is shown as a
solid line in the left panel of Fig.~\ref{constraints}, while the
shaded region is the union of the inequalities (\ref{mphi_max}) and
(\ref{mphi_min}).

In the regime of strong self-interaction $\Lambda_{3} \gg 1$, on the
other hand, we employ a technique similar to the one discussed above
to obtain again upper and lower limits, with the difference that in this
regime there is only a single mass curve (\cf Fig.~\ref{lambdabig}).
In this case, the maximum $m_\phi$ is
\begin{equation}
\label{large_lambda_SgrA_max}
m_{\phi} \leq m^{\rm max}_{\phi}=
\sqrt{\frac{0.22\, m_{_{P}}^3}{\Msgr}}
\lambda^{1/4} \simeq 2.9\times 10^5 \,\lambda^{1/4}~\mbox{eV}\,.
\end{equation}
The condition (\ref{large_lambda_SgrA_max}) when the equality holds is
shown as a dashed line in the right panel of Fig.~\ref{constraints}.
In addition, also when $\Lambda_3 \gg 1$, $C_{\rm min}$ allows us to
derive the lower limit of $m_\phi$ with the difference that the
variable parameter now is just ${\tilde B}_{\rm c}$ and we do not need
to iterate also on $\Lambda$. As a result we obtain
\begin{equation}
\label{large_lambda_SgrA_min}
m_{\phi} \geq m^{\rm min}_{\phi} \simeq 
3.7 \times 10^4 \,\lambda^{1/4}~\mbox{eV}\,.
\end{equation}
The condition (\ref{mphi_min}) when the equality holds is shown as a
solid line in the right panel of Fig.~\ref{constraints}, while the
shaded region is the union of the inequalities (\ref{mphi_max}) and
(\ref{large_lambda_SgrA_min}). The overlap in the constraints between
the two regimes is very good and hence the results obtained for
$\Lambda_3 \gg 1$ can be easily extrapolated to the regime $\Lambda_3
\ll 1$ by using fitting expressions.

\subsection{Constraints from NGC 4258}

The procedure discussed above for the Galactic centre can be exploited
in principle for any other galaxy for which there is an accurate
measurement of the mass and dimensions of the central object. As an
example, we here consider also NGC 4258, which is a spiral galaxy at a
distance of about $73-83$ Mpc. In this case, the mass of the central
DCO has been estimated to be $M_{\text{NGC 4258}}=38.1\pm0.01 \times
10^6\, M_\odot$, while the observations of the rotation curve require
a central density of at least $4 \times 10^9\, M_\odot\,
\mbox{pc}^{-3}$, which implies a maximum radius $R_{\rm max}\simeq
36000\,R_{\rm S}$~\cite{Herrnstein:2005xc}.

Restricting our attention to the regime $\Lambda_{3} \ll 1$, we show
in Fig.~\ref{comparaNGC-SgrA} as a forward-shaded area the upper and
lower limits for $(m_\phi,\lambda)$ as derived from the observations
coming from NGC 4258. Similarly, in the regime $\Lambda_{3} \gg 1$,
the upper limit for the scalar-field mass is obtained by setting the
maximum mass \mbox{$m_{\phi} \leq m^{\rm max}_{\phi} = \sqrt{{0.22\,
      m_{_{P}}^3}/M_{\text{NGC 4258}}}\lambda^{1/4}$}, while the lower
limit comes from the minimum compactness $C_{\rm min}=1/72000$. The
combined constraint can then be expressed as (not shown in
Fig.~\ref{comparaNGC-SgrA})
\begin{equation}\label{large_lambda_NGC}
6.3~\mbox{eV} 
\lesssim \frac{m_\phi}{\lambda^{1/4}} \lesssim 
9.6\times 10^4~\mbox{eV} \,.
\end{equation}

Combining these constraints with those derived in the
previous Section for $\sgr$, we obtain the following global
constraints on the parameter space obtained when modelling the DCOs as
BSs
\begin{equation}\label{overlap_large_lambda}
3.7 \times 10^4~\mbox{eV}
\lesssim  \frac{m_\phi}{\lambda^{1/4}} \lesssim
 9.6\times 10^4~\mbox{eV}\,.
\end{equation}
The corresponding region is shown as a double shaded region in
Fig.~\ref{comparaNGC-SgrA} and clearly sets a tighter constraint on
the possible values of $m_{\phi}$ and $\lambda$.

\FIGURE[t]{ 
\centerline{
\includegraphics[width=8.0cm]{./figs/fig6}}
\caption{ Constraints for $m_{\phi}$ and $\lambda$ derived from the
  observations of NGC 4258 in the limit $\Lambda_{3} \ll 1$
  (forward-shaded area). Shown as a double-shaded area is the
  combination of the constraints coming from $\sgr$ and NGC
  4258. Shown with the inset is the same data but in terms of the
  dimensionless self-interacting coupling constant $\Lambda$.}
\label{comparaNGC-SgrA}
}

\subsection{Constraints from Dark-Matter Models}
\label{DMmodels}

As mentioned in the Introduction, constraints on the properties of the
scalar field can be imposed also exploiting cosmological
observations. We recall that the Lambda Cold Dark Matter model (or
$\Lambda$CDM as it is usually referred to\footnote{Note that the
  $\Lambda$ here is the cosmological constant and should not be
  confused with the self-coupling constant defined
  in~(\ref{Lambda}).})  is a standard model of big bang cosmology,
which attempts to explain within a single framework a number of
cosmological constraints and observations. These are, for instance,
the existence and properties of the cosmic microwave background, the
abundances of light elements (hydrogen, helium, lithium, the
large-scale structure of the universe in terms of galaxy clusters,
and, more recently, also the accelerating expansion of the universe
observed in the light from distant galaxies and supernovae.

In spite of its many successes, some of the predictions of the
$\Lambda$CDM scenario are in contrast with observations. One example
is the so-called the ``core/cusp problem''~\cite{Navarro:1996gj,
  Gentile:2004tb}, where numerical N-body simulations assuming a
$\Lambda$CDM cosmology predict the generation of diverging cusps in
the central density of the DM distribution in galaxies, and which are
incompatible with observations at galactic
scales~\cite{Salucci:2007}. Another example is the so-called the
``dwarf-galaxy problem''~\cite{Klypin:1999uc}, where again the
$\Lambda$CDM predicts an overproduction of dwarf galaxies in the local
group, which is not corroborated by observations. Some possible
solutions to these problems have been suggested even within the
$\Lambda$CDM paradigm. For instance, the core/cusp problem could be
alleviated if dynamical friction~\cite{ElZant:2001re} or stellar
feedback~\cite{Mashchenko:2006dm} are included. Similarly, for the
dwarf-galaxy problem, it was suggested that accretion of gas onto
low-mass halos is not enough to make them observable. Possible reasons
of this inefficient accretion could be supernova
feedback~\cite{Dekel:1986gu,Ferrara:2000}, 
photoionization~\cite{Efstathiou1992} or
reionization~\cite{Bullock:2000wn}. It has also been suggested that
this may simply be a 
problem with our ability to detect the
missing satellite galaxies in the Local Group, or that these
are in a region of the sky that has not been surveyed yet~\cite{Bullock:2010uy}.
Alternative solutions suggest instead that DM might be treated as
extremely light bosonic (dark) matter. Some models based on this idea
are the so-called ``fuzzy dark matter''~\cite{Hu:2000ke}, or the
scalar-field (dark) matter~\cite{sflocal2000, sflocal2001,
  sflocal2004}. Both suggestions predict flat density profiles at the
centre of galaxies~\cite{Bernal08} and fit the abundance of dwarf
galaxies~\cite{sflocal2000,sflocal2001,sflocal2004}. An alternative
suggestion is that the DM particle could have a very strong
self-interaction, but negligible
annihilation~\cite{Spergel:1999mh}. In this case, the core/cusp
problem is alleviated since the strong self-interaction increases the
scattering among DM particles. The scattering cross section is then so
high that collisions at the centre of galaxies are very so frequent
that dark matter particles scatter out the centre as fast as they are
accreted, thus effectively preventing a growth in the density.

In what follows we briefly summarize the main results of the
scalar-field DM and the ``collisional'' DM scenarios and discuss how
to use the present observations to further constraint the space of
parameters for a complex scalar field. We recall that the collisional
DM model of~\cite{Spergel:1999mh} is also referred to as
``self-interacting'' DM because the scattering cross section is much
larger than that of standard WIMPS, that collisions among DM particles
are very frequent.  On the other hand, the prototype of a DM candidate
consisting of ultra-light scalar field is the axion, which has been
introduced as a solution to the strong CP problem~\cite{Peccei:1977hh,
  Peccei:1977ur}. The mass of the axion has been estimated to be in
the range $10^{-5}~\mbox{eV}\lesssim m_a \lesssim 10^{-3}\,{\rm
  eV}$~\cite{Raffelt:2006cw}. However, in order to explain the
singular galactic core and suppress the formation of dwarf galaxies,
the axion's mass should be orders of magnitude smaller. In the absence
of self-interaction, the mass should in fact be $m_\phi \sim
10^{-22}-10^{-24}\,{\rm eV}$, which then gives a reasonable fit to the
rotation curves~\cite{arbey1,arbey2}.

At the same time, the cosmological study of the scalar-field DM
performed in~\cite{sflocal2000, Matos:2009hf, Matos:2008ag} for
different potentials $V(\phi)$ have shown to be able to reproduce all
the features of the standard $\Lambda$CDM in the linear regime of
perturbations. Furthermore, a DM model with a complex scalar field
having a quartic potential was studied in~\cite{Arbey:2003sj}, and the
fit of rotational curves for dwarf galaxies was shown to be robust
provided that
\begin{equation}
\label{arbey_limit}
2.7 ~{\rm eV}
\lesssim \frac{m_\phi}{\lambda^{1/4}} \lesssim
2.9 ~{\rm eV}\,.
\end{equation}
This is shown as a thick solid line in Fig.~\ref{DM_constraints}.

A different suggestion has been made in Ref.~\cite{Spergel:1999mh},
where the DM is proposed to be cold, non-dissipative and
self-interacting with a scattering cross-section per particle mass
given by
\begin{equation}
\label{collisional_strenght}
\frac{\sigma_{2 \to 2}}{m_\phi}=10^{-25}-10^{-23}~\frac{\mbox{cm}^2}{\mbox{GeV}}\,,
\end{equation}
and independent of the particle itself. This model solves the small
scale problems of $\Lambda$CDM and $N-$body numerical simulation
confirms this~\cite{Dave:2000ar}.  Since the conjecture is made
without assuming a particular model for DM, we can derive the values
for $(m_\phi,\lambda)$ for a scalar field with self-interaction. For
that, we introduce as the interacting potential
$V={\lambda}\phi^4/4$. Thus, the scattering matrix element is
$\mathcal{M(\phi\phi \to \phi\phi)}=i\lambda$ and the resulting
cross-section is
\begin{equation}\label{cs_scalar}
\sigma(\phi\phi \to \phi\phi)=
\sigma_{\phi \phi}=\frac{\lambda^2}{16 \pi s}=
\frac{\lambda^2}{64 \pi m_\phi^2}\,,
\end{equation}
since the square of the centre of mass energy is $s=4 m_\phi^2$. By
requiring that the cross-section of Eq.~(\ref{cs_scalar}) has the
strength of the collisional DM of Eq.~(\ref{collisional_strenght}), we
find that $m_\phi$ must be
\begin{equation}
9.5 \times 10^{5}~{\rm eV}
\lesssim \frac{m_\phi}{\lambda^{2/3}} \lesssim
9.5 \times 10^7~{\rm eV}\,.
\end{equation}
The constraints coming from the different DM models are summarized in
Fig.~\ref{DM_constraints}, where we show with a dashed magenta area
the allowed values for $m_{\phi}$ and $\lambda$ coming from the
collisional DM model [\cf eq.~(\ref{collisional_strenght})], while
shown with a black region are the corresponding values as constrained
from the measurement of the DM halo [\cf eq.~(\ref{arbey_limit})].

\FIGURE[t]{ 
\centerline{
\includegraphics[width=8.0cm]{./figs/DM_constraints}}
\caption{The constraints on $m_{\phi}$ and $\lambda$ coming from the
  different DM models. Shown with a dashed magenta area the allowed
  values for coming from the collisional DM model, while shown with a
  black region are the corresponding values as constrained from the
  measurement of the DM halo.}
\label{DM_constraints}
}

\FIGURE[t]{ 
\centerline{
    \includegraphics[width=14.0cm]{./figs/globalnew}}
\caption{Constraints on $(m_\phi,\lambda)$ based on available stellar
  kinematic data for $\sgr$ (narrow upper stripe filled with solid
  blue lines), as well as for NGC 4258 (broad lower stripe filled with
  dashed orange lines) and their intersection (middle stripe filled
  with black crosses). The solid magenta stripe that crosses all other
  stripes is derived from the requirement of collisional DM and
  demarcates the combined restrictions (red solid region).  We also
  show the limits on the parameters derived from the models of
  galactic DM halos based on scalar fields (solid black line). }
\label{global}
}

\subsection{Collecting all constraints}

We can combine all of the constraints derived so far and conclude that
a scalar field which can account for DM candidate and fulfill the
constraints imposed by the observations on the DCO at the centre of
our Galaxy and NGC 4258 must have a mass and a self-interacting
coupling constant in the ranges
\begin{eqnarray}
6.5 \times 10^{-9} & \lesssim & \lambda \lesssim 4.2 \times 10^{-3}\,,
\nonumber \\ 
332~\mbox{eV}     & \lesssim & m_\phi \lesssim 2.46 \times 10^4 ~\mbox{eV}\,.
\end{eqnarray}
This region is marked as ``combined'' in Fig.~\ref{global}, which
reports also the constraints coming from modelling the DCOs as BSs and
discussed in the previous Sections.

We also note that although the lower limit of the region constrained
from the observations of NGC 4258 is not too far from  the limit
set by condition~(\ref{arbey_limit}), the latter is at least four
orders of magnitude smaller than the region representing the overlap
between collisional DM models and the kinematic constraints from
galactic observations. As a result, we conclude that a scalar field
that could explain the rotation curves in nearby galaxies cannot be
the same composing a BS representing the DCOs in galactic nuclei.

\section{Conclusions}\label{conclusions}

We have used the most accurate astronomical observations on the
kinematics of galaxy centres to set constraints on the mass $m_{\phi}$
and self-interaction constant $\lambda$ of complex scalar fields,
which have been employed in the literature to model BSs as DCOs in
galactic centres or as DM candidates on cosmological scales. More
specifically, we have used the estimates of the mass and volume for
the nucleus of our Galaxy and of NGC 4258 to limit the possible
models of BSs when these are advocated to account for such
observations.

These constraints, together with those obtained by considering
possible DM candidates have allowed us to restrict the possible values
of $m_{\phi}$ and $\lambda$ to the ranges
\begin{eqnarray}
6.5 \times 10^{-9} & \lesssim & \lambda \lesssim 4.2 \times 10^{-3}\,,
\nonumber \\ 
332~\mbox{eV}     & \lesssim & m_\phi \lesssim 2.46 \times 10^4 ~\mbox{eV}\,,
\nonumber 
\end{eqnarray}
that is to a region that spans six orders of magnitude for $\lambda$
and one order of magnitude for $m_\phi$. To the best of our knowledge,
these constraints for the scalar field are by far tighter than any
other discussed so far in astrophysical or cosmological context. In
addition, since the space of parameters has two non-overlapping
constrained regions, we conclude that a scalar field that could
explain the rotation curves in nearby galaxies cannot be the same
composing a BS representing the DCOs in galactic nuclei.

Our analysis here has been limited to the simplest possible scenario
in which the DCO is not a black hole but a BS. However, as long as the
presence of a BS at the galactic centre cannot be ruled out, a more
plausible configuration is one in which the BS actually hosts at its
centre a massive black hole of comparable mass, \ie a {\em hybrid BS}.
Mixed models for BS have been proposed in the past, for instance when
studying the boundary between stable and unstable equilibrium
configurations of cold boson-fermion stars
(see~\cite{Jetzer90,Jetzer92} and references therein).

While we are not suggesting that a hybrid BS hosting a BH at its
centre is the most natural configuration, we do believe it is more
appealing than a simple BS and for a number of reasons. First, as long
as the size of the black-hole horizon is much smaller than the
scalar-field Compton length, the accretion of the scalar field onto
the black hole is very small~\cite{UrenaLopez:2002du} and hence a
stationary model can be constructed (if this was not the case the
scalar field would eventually be all accreted by the black hole, as
recently suggested in~\cite{CruzOsorio:2010qs}). Second, a hybrid BS
automatically satisfies all the constraints usually set when
considering the DCO as a massive black hole. Finally, by having a
black hole at its centre, a hybrid BS does not need to account for the
electromagnetic emission that would be otherwise produced by the
matter condensing and heating-up at the centre of an ordinary BS.

Together with these advantages, however, a hybrid-BS model has the
important drawback that it may be very difficult to distinguish it
from a pure black-hole solution, the differences being so small that
they may be below the present (and possibly future) observational
limits from electromagnetic radiation. In this respect, the
gravitational-wave observations that will be made possible by the
space-borne LISA mission, may well be determinant. LISA will in fact
detect a large number of extreme-mass ratio inspirals or EMRIs
(see,~\cite{Amaro-SeoaneEtAl07} for a recent review) and because the
associated gravitational waveforms probe a region of the spacetime
very close to the black hole, they may work as telltale about the
presence of a scalar field surrounding the black hole. Some work on
the effects of matter fields on the gravitational-wave emission from
EMRIs has already been done~\cite{Barausse:2007, Barausse:2008}, and
extending this question also to scalar fields will be the focus of our
future research.

\section{Acknowledgments}

It is a pleasure to thank Marc Freitag for useful discussions and Rainer
Sch{\"o}del for input on the observational aspects. We are grateful to Monica
Colpi, Francisco Guzman, Diego Torres, Philippe Jetzer, Paolo Salucci, and
David J. Vanecek for their comments on the manuscript. This work was supported
in part by the DFG grant SFB/Transregio~7 ``Gravitational-Wave Astronomy''.

\providecommand{\href}[2]{#2}\begingroup\raggedright\endgroup


\begin{thebibliography}{10}

\bibitem{Bertone:2004pz}
G.~{Bertone}, D.~{Hooper}, and J.~{Silk}, {\it {Particle dark matter: evidence,
  candidates and constraints}},  {\em Phys. Rept.} {\bf 405} (2005)
  279--390

\bibitem{Hu:2000ke}
W.~{Hu}, R.~{Barkana}, and A.~{Gruzinov}, {\it {Fuzzy Cold Dark Matter: The
  Wave Properties of Ultralight Particles}},  {\em Phys. Rev. Lett.}
  {\bf 85} (2000) 1158--1161

\bibitem{sflocal2000} T.~{Matos} and L.~A. {Ure{\~n}a-L{\'o}pez}, {\it
  Quintessence and scalar dark matter in the Universe}, {\em Class. and Quant. Grav.} 
  {\bf 17} (2000) L75--L81

\bibitem{sflocal2001}
T.~{Matos} and L.~{Arturo Ure{\~n}a-L{\'o}pez}, {\it {Further analysis of a
  cosmological model with quintessence and scalar dark matter}},  {\em Phys. Rev. D} 
  {\bf 63} (2001) 063506

\bibitem{sflocal2004}
T.~{Matos} and L.~A. {Ure{\~n}a-L{\'o}pez}, {\it {On the Nature of Dark
  Matter}},  {\em Int. J. Mod. Phys. D} {\bf 13} (2004)
  2287--2291

\bibitem{Frieman:1995pm}
J.~A. {Frieman}, C.~T. {Hill}, A.~{Stebbins}, and I.~{Waga}, {\it {Cosmology
  with Ultralight Pseudo Nambu-Goldstone Bosons}},  {\em Phys. Rev.
  Lett.} {\bf 75} (1995) 2077--2080

\bibitem{Guth:1980zm}
A.~H. {Guth}, {\it {Inflationary universe: A possible solution to the horizon
  and flatness problems}},  {\em Phys. Rev. D} {\bf 23} (1981) 347--356.

\bibitem{Linde:1981mu}
A.~D. {Linde}, {\it {A new inflationary universe scenario: A possible solution
  of the horizon, flatness, homogeneity, isotropy and primordial monopole
  problems}},  {\em Phys. Lett. B} {\bf 108} (1982) 389--393.

\bibitem{Albrecht:1982mp}
A.~{Albrecht}, P.~J. {Steinhardt}, M.~S. {Turner}, and F.~{Wilczek}, {\it
  {Reheating an inflationary universe}},  {\em Phys. Rev. Lett.} {\bf
  48} (1982) 1437--1440.

\bibitem{Lee:1986ts}
T.~D. {Lee}, {\it {Soliton stars and the critical masses of black holes}},
  {\em Phys. Rev. D} {\bf 35} (1987) 3637--3639.

\bibitem{Seidel:1991zh} E.~Seidel and W.~M.~Suen, {\it Oscillating
  soliton stars}, {\em Phys.\ Rev.\ Lett.}  {\bf 66} (1991) 1659.

\bibitem{UrenaLopez:2001tw} L.~A. {Ure{\~n}a-L{\'o}pez}, {\it
  {Oscillatons revisited}}, {\em Class. and Quant. Grav.} {\bf 19}
  (2002) 2617--2632

\bibitem{Alcubierre:2003sx} M.~Alcubierre, R.~Becerril, S.~F.~Guzman,
  T.~Matos, D.~Nunez and L.~A.~Urena-Lopez, {\it Numerical studies of
    phi**2-oscillatons}, {\em Class.\ Quant.\ Grav.}  {\bf 20} (2003)
  2883.

\bibitem{Jetzer92} P.~{Jetzer}, {\it {Boson stars}}, {\em Phys. Rept.}
  {\bf 220} (1992) 163--227.

\bibitem{Schunck:2003kk}
F.~E. {Schunck} and E.~W. {Mielke}, {\it {TOPICAL REVIEW: General relativistic
  boson stars}},  {\em Class. and Quant. Grav.} {\bf 20} (2003)
  301.

\bibitem{arbey1}
A.~{Arbey}, J.~{Lesgourgues}, and P.~{Salati}, {\it {Quintessential halos
  around galaxies}},  {\em Phys. Rev. D} {\bf 64} (2001) 123528

\bibitem{arbey2}
A.~{Arbey}, J.~{Lesgourgues}, and P.~{Salati}, {\it {Cosmological constraints
  on quintessential halos}},  {\em Phys. Rev. D} {\bf 65} (2002)
  083514

\bibitem{Matos:2007zza}
T.~{Matos} and L.~A. {Ure{\~n}a-L{\'o}pez}, {\it {Flat rotation curves in
  scalar field galaxy halos}},  {\em Gen. Rel. Grav.} {\bf
  39} (2007) 1279--1286.

\bibitem{Arbey:2003sj}
A.~{Arbey}, J.~{Lesgourgues}, and P.~{Salati}, {\it {Galactic halos of fluid
  dark matter}},  {\em Phys. Rev. D} {\bf 68} (2003) 023511

\bibitem{HernandezEtAl04}
X.~{Hern{\'a}ndez}, T.~{Matos}, R.~A. {Sussman}, and Y.~{Verbin}, {\it {Scalar
  field ``mini-MACHOs'': A new explanation for galactic dark matter}},  {\em
 Phys. Rev. D} {\bf 70} (2004) 043537

\bibitem{Barranco:2010ib}
J.~{Barranco} and A.~{Bernal}, {\it {Self-gravitating system made of axions}},
  {\em ArXiv e-prints} (2010)
  [\href{http://xxx.lanl.gov/abs/1001.1769}{{\tt arXiv:1001.1769}}].

\bibitem{Torres:2000dw}
D.~F. {Torres}, S.~{Capozziello}, and G.~{Lambiase}, {\it {Supermassive boson
  star at the galactic center?}},  {\em Phys. Rev. D} {\bf 62} (2000)
  104012

\bibitem{Guzman:2005bs}
F.~S. {Guzm{\'a}n}, {\it {Accretion disk onto boson stars: A way to supplant
  black hole candidates}},  {\em Phys. Rev. D} {\bf 73} (2006) 021501

\bibitem{Guzman:2009zz}
F.~S. {Guzm{\'a}n} and J.~M. {Rueda-Becerril}, {\it {Spherical boson stars as
  black hole mimickers}},  {\em Phys. Rev. D} {\bf 80} (2009) 084023.

\bibitem{GhezEtAl03}
A.~M. {Ghez}, S.~{Salim}, S.~D. {Hornstein}, A.~{Tanner}, J.~R. {Lu},
  M.~{Morris}, E.~E. {Becklin}, and G.~{Duch{\^ e}ne}, {\it {Stellar Orbits
  around the Galactic Center Black Hole}},  {\em Astrophys. J.} {\bf 620} (2005)
  744--757.

\bibitem{SchoedelEtAl03}
R.~{Sch{\"o}del}, T.~{Ott}, R.~{Genzel}, A.~{Eckart}, N.~{Mouawad}, and
  T.~{Alexander}, {\it {Stellar Dynamics in the Central Arcsecond of Our
  Galaxy}},  {\em Astrophys. J.} {\bf 596} (2003) 1015--1034.

\bibitem{FF04} L.~{Ferrarese} and H.~{Ford}, {\it {Supermassive Black Holes in
Galactic Nuclei: Past, Present and Future Research}}, {\em Space Science
Reviews} {\bf 116} (2005) 523--624.


\bibitem{MoranEtAl99}
J.~M. {Moran}, L.~J. {Greenhill}, and J.~R. {Herrnstein}, {\it {Observational
  Evidence for Massive Black Holes in the Centers of Active Galaxies}},  {\em
  Journal of Astrophysics and Astronomy} {\bf 20} (1999) 165.

\bibitem{Kormendy03}
J.~{Kormendy}, {\it {The Stellar-Dynamical Search for Supermassive Black Holes
  in Galactic Nuclei}},  in {\em ``Coevolution of Black Holes and Galaxies'',
  Carnegie Observatories, Pasadena} (L.~{Ho}, ed.), 2003.

\bibitem{GhezEtAl03b}
A.~M. {Ghez}, G.~{Duch{\^e}ne}, K.~{Matthews}, S.~D. {Hornstein}, A.~{Tanner},
  J.~{Larkin}, M.~{Morris}, E.~E. {Becklin}, S.~{Salim}, T.~{Kremenek},
  D.~{Thompson}, B.~T. {Soifer}, G.~{Neugebauer}, and I.~{McLean}, {\it {The
  First Measurement of Spectral Lines in a Short-Period Star Bound to the
  Galaxy's Central Black Hole: A Paradox of Youth}},  {\em Astrophys. J. Lett.} {\bf 586}
  (2003) L127--L131.

\bibitem{EisenhauerEtAl05} F.~{Eisenhauer}, R.~{Genzel},
  T.~{Alexander}, R.~{Abuter}, T.~{Paumard}, T.~{Ott}, A.~{Gilbert},
  S.~{Gillessen}, M.~{Horrobin}, S.~{Trippe}, H.~{Bonnet}, C.~{Dumas},
  N.~{Hubin}, A.~{Kaufer}, M.~{Kissler-Patig}, G.~{Monnet},
  S.~{Str{\"o}bele}, T.~{Szeifert}, A.~{Eckart}, R.~{Sch{\"o}del}, and
  S.~{Zucker}, {\it {SINFONI in the Galactic Center: Young Stars and
      Infrared Flares in the Central Light-Month}}, {\em
    Astrophys. J.} {\bf 628} (2005) 246--259.

\bibitem{GhezEtAl08}
A.~M. {Ghez}, S.~{Salim}, N.~N. {Weinberg}, J.~R. {Lu}, T.~{Do}, J.~K. {Dunn},
  K.~{Matthews}, M.~R. {Morris}, S.~{Yelda}, E.~E. {Becklin}, T.~{Kremenek},
  M.~{Milosavljevic}, and J.~{Naiman}, {\it {Measuring Distance and Properties
  of the Milky Way's Central Supermassive Black Hole with Stellar Orbits}},
  {\em Astrophys. J.} {\bf 689} (2008) 1044--1062.

\bibitem{GillessenEtAl09} S.~{Gillessen}, F.~{Eisenhauer},
  S.~{Trippe}, T.~{Alexander}, R.~{Genzel}, F.~{Martins}, and
  T.~{Ott}, {\it {Monitoring Stellar Orbits Around the Massive Black
      Hole in the Galactic Center}}, {\em Astrophys. J.} {\bf 692}
  (2009) 1075--1109.

\bibitem{Kaup:1968zz}
D.~J. {Kaup}, {\it {Klein-Gordon Geon}},  {\em Phys. Rev.} {\bf 172}
  (1968) 1331--1342.

\bibitem{Ruffini:1969qy}
R.~{Ruffini} and S.~{Bonazzola}, {\it {Systems of Self-Gravitating Particles in
  General Relativity and the Concept of an Equation of State}},  {\em Phys.
  Rev.} {\bf 187} (1969) 1767--1783.

\bibitem{Colpi:1986ye}
M.~{Colpi}, S.~L. {Shapiro}, and I.~{Wasserman}, {\it {Boson stars -
  Gravitational equilibria of self-interacting scalar fields}},  {\em Phys.
  Rev. Lett.} {\bf 57} (1986) 2485--2488.

\bibitem{SeidelSuen90}
E.~{Seidel} and W.~{Suen}, {\it {Dynamical evolution of boson stars: Perturbing
  the ground state}},  {\em Phys. Rev. D} {\bf 42} (1990) 384--403.

\bibitem{HawleyChoptuik00}
S.~H. {Hawley} and M.~W. {Choptuik}, {\it {Boson stars driven to the brink of
  black hole formation}},  {\em Phys. Rev. D} {\bf 62} (2000) 104024.

\bibitem{Guzman:2004jw} F.~S.~Guzman, {\it Evolving spherical boson
  stars on a 3D cartesian grid}, {\em Phys.\ Rev.\ D }{\bf 70} (2004)
  044033



\bibitem{GleiserWatkins89}
M.~{Gleiser} and R.~{Watkins}, {\it {Gravitational stability of scalar
  matter}},  {\em Nucl. Phys. B} {\bf 319} (1989) 733--746.

\bibitem{LeePang89}
T.~D. {Lee} and Y.~{Pang}, {\it {Stability of mini-boson stars}},  {\em Nucl.
  Phys. B} {\bf 315} (1989) 477--516.

\bibitem{KusmartsevMielkeSchunck91}
F.~V. {Kusmartsev}, E.~W. {Mielke}, and F.~E. {Schunck}, {\it {Gravitational
  stability of boson stars}},  {\em Phys. Rev. D} {\bf 43} (1991)
  3895--3901.

\bibitem{Guzman09} F.~S.~Guzman,, {\it {The three dynamical fates of
    Boson Stars}}, {\em Revista Mexicana de F{\'\i}sica} {\bf 55}
  (2009) 321--326.  

\bibitem{Buchdahl:59}
H.~A. {Buchdahl}, {\it {General Relativistic Fluid Spheres}},  {\em Phys.
  Rev.} {\bf 116} (1959) 1027--1034.

\bibitem{GenzelEtAl10}
R.~{Genzel}, F.~{Eisenhauer}, and S.~{Gillessen}, {\it {The Massive Black Hole
  and Nuclear Star Cluster in the Center of the Milky Way}},  {\em ArXiv
  e-prints} (2010) [\href{http://xxx.lanl.gov/abs/1006.0064}{{\tt
  arXiv:1006.0064}}].

\bibitem{DoelemanEtAl08}
S.~S. {Doeleman}, J.~{Weintroub}, A.~E.~E. {Rogers}, R.~{Plambeck},
  R.~{Freund}, R.~P.~J. {Tilanus}, P.~{Friberg}, L.~M. {Ziurys}, J.~M. {Moran},
  B.~{Corey}, K.~H. {Young}, D.~L. {Smythe}, M.~{Titus}, D.~P. {Marrone}, R.~J.
  {Cappallo}, D.~{Bock}, G.~C. {Bower}, R.~{Chamberlin}, G.~R. {Davis}, T.~P.
  {Krichbaum}, J.~{Lamb}, H.~{Maness}, A.~E. {Niell}, A.~{Roy},
  P.~{Strittmatter}, D.~{Werthimer}, A.~R. {Whitney}, and D.~{Woody}, {\it
  {Event-horizon-scale structure in the supermassive black hole candidate at
  the Galactic Centre}},  {\em Nature} {\bf 455} (2008) 78--80.

\bibitem{DoelemanEtAl09}
S.~{Doeleman}, E.~{Agol}, D.~{Backer}, F.~{Baganoff}, G.~C. {Bower},
  A.~{Broderick}, A.~{Fabian}, V.~{Fish}, C.~{Gammie}, P.~{Ho}, M.~{Honman},
  T.~{Krichbaum}, A.~{Loeb}, D.~{Marrone}, M.~{Reid}, A.~{Rogers},
  I.~{Shapiro}, P.~{Strittmatter}, R.~{Tilanus}, J.~{Weintroub}, A.~{Whitney},
  M.~{Wright}, and L.~{Ziurys}, {\it {Imaging an Event Horizon: submm-VLBI of a
  Super Massive Black Hole}},  in {\em astro2010: The Astronomy and
  Astrophysics Decadal Survey}, vol.~2010 of {\em ArXiv Astrophysics e-prints},
  pp.~68, 2009.
\newblock \href{http://xxx.lanl.gov/abs/0906.3899}{{\tt arXiv:0906.3899}}.

\bibitem{Herrnstein:2005xc}
J.~R. {Herrnstein}, J.~M. {Moran}, L.~J. {Greenhill}, and A.~S. {Trotter}, {\it
  {The Geometry of and Mass Accretion Rate through the Maser Accretion Disk in
  NGC 4258}},  {\em Astrophys. J.} {\bf 629} (2005) 719--738.

\bibitem{Navarro:1996gj}
J.~F. {Navarro}, C.~S. {Frenk}, and S.~D.~M. {White}, {\it {A Universal Density
  Profile from Hierarchical Clustering}},  {\em Astrophys. J.} {\bf 490} (1997)
  493.

\bibitem{Gentile:2004tb} G.~Gentile, P.~Salucci, U.~Klein, D.~Vergani
  and P.~Kalberla, {\it The cored distribution of dark matter in
    spiral galaxies}, {\em Mon.\ Not.\ Roy.\ Astron.\ Soc.} {\bf 351},
    (2004), 903


\bibitem{Salucci:2007} P. Salucci, A. Lapi, C. Tonini, G. Gentile,
  I. Yegorova, and U. Klein, U.  {\em Mon.\ Not.\ Roy.\ Astron.\ Soc.}
  {\bf 378}, (2007), 41

\bibitem{Klypin:1999uc} A.~{Klypin}, A.~V. {Kravtsov},
  O.~{Valenzuela}, and F.~{Prada}, {\it {Where Are the Missing
      Galactic Satellites?}}, {\em Astrophys. J.} {\bf 522} (1999)
  82--92.

\bibitem{ElZant:2001re} A.~El-Zant, I.~Shlosman and Y.~Hoffman, ``Dark
  Halos: The Flattening of the Density Cusp by Dynamical Friction,''
  arXiv:astro-ph/0103386.

\bibitem{Mashchenko:2006dm} S.~Mashchenko, H.~M.~P.~Couchman and
  J.~Wadsley, {\it Cosmological puzzle resolved by stellar feedback in
    high redshift galaxies}, {\em Nature} {\bf 442} (2006) 539.


\bibitem{Dekel:1986gu} A.~Dekel and J.~Silk, {\it The origin of dwarf
  galaxies, cold dark matter, and biased galaxy formation}, {\em
  Astrophys.\ J.}  {\bf 303} (1986) 39.

\bibitem{Ferrara:2000}  A. ~Ferrara and E. ~Tolstoy {\it The role of 
stellar feedback and dark matter in the evolution of dwarf galaxies}, 
{\em Monthly Notices of the Royal Astronomical Society} {\bf 313} (2000)
  291-313.


\bibitem{Efstathiou1992}  G. ~Efstathiou, {\it Suppressing the formation 
of dwarf galaxies via photoionization}, {\em Monthly
    Notices of the Royal Astronomical Society} {\bf 256} (1992)
  43P-47P.

\bibitem{Bullock:2000wn} J.~S.~Bullock, A.~V.~Kravtsov and
  D.~H.~Weinberg, `Reionization and the abundance of galactic
  satellites,'' Astrophys.\ J.\ {\bf 539}, 517 (2000)
  [arXiv:astro-ph/0002214].

\bibitem{Bullock:2010uy}
  J.~S.~Bullock,
  {\it Notes on the Missing Satellites Problem},
  {\em ArXiv e-prints} (2010)
  [\href{http://xxx.lanl.gov/abs/1009.4505}{{\tt arXiv:1009.4505}}]

\bibitem{Bernal08} A.~{Bernal}, T.~{Matos}, and D.~{N{\'u}{\~n}ez},
  {\it {Flat Central Density Profiles from Scalar Field Dark Matter
      Halos}}, {\em Revista Mexicana de Astronom{\'\i}a y
    Astrof{\'\i}sica } {\bf 44} (2008) 149--160.

\bibitem{Spergel:1999mh} D.~N. {Spergel} and P.~J. {Steinhardt}, {\it
  {Observational Evidence for Self-Interacting Cold Dark Matter}},
  {\em Phys. Rev. Lett.} {\bf 84} (2000) 3760--3763.

\bibitem{Peccei:1977hh}
R.~D. {Peccei} and H.~R. {Quinn}, {\it {CP conservation in the presence of
  pseudoparticles}},  {\em Phys. Rev. Lett.} {\bf 38} (1977)
  1440--1443.

\bibitem{Peccei:1977ur} R.~D. {Peccei} and H.~R. {Quinn}, {\it
  {Constraints imposed by CP conservation in the presence of
    pseudoparticles}}, {\em Phys. Rev. D} {\bf 16} (1977) 1791--1797.

\bibitem{Raffelt:2006cw}
G.~G. {Raffelt}, {\it {Astrophysical Axion Bounds}},  in {\em Axions}
  ({M.~Kuster, G.~Raffelt, \& B.~Beltr{\'a}n}, ed.), vol.~741 of {\em Lecture
  Notes in Physics, Berlin Springer Verlag}, pp.~51, 2008.


\bibitem{Matos:2009hf}
T.~{Matos}, J.~{Lu{\'e}vano}, I.~{Quiros}, L.~A. {Ure{\~n}a-L{\'o}pez}, and
  J.~A. {V{\'a}zquez}, {\it {Dynamics of scalar field dark matter with a
  cosh-like potential}},  {\em Phys. Rev. D} {\bf 80} (2009) 123521.

\bibitem{Matos:2008ag} T.~{Matos}, A.~{V{\'a}zquez-Gonz{\'a}lez}, and
  J.~{Maga{\~n}a}, {\it {Phi square as dark matter}}, {\em Monthly
    Notices of the Royal Astronomical Society} {\bf 393} (2009)
  1359--1369.

\bibitem{Dave:2000ar}
R.~{Dav{\'e}}, D.~N. {Spergel}, P.~J. {Steinhardt}, and B.~D. {Wandelt}, {\it
  {Halo Properties in Cosmological Simulations of Self-interacting Cold Dark
  Matter}},  {\em Astrophys. J.} {\bf 547} (2001) 574--589.

\bibitem{Jetzer90}
P.~{Jetzer}, {\it {Stability of combined boson-fermion stars}},  {\em Phys.
  Lett. B} {\bf 243} (1990) 36--40.

\bibitem{UrenaLopez:2002du}
L.~A. {Ure{\~n}a-L{\'o}pez} and A.~R. {Liddle}, {\it {Supermassive black holes
  in scalar field galaxy halos}},  {\em Phys. Rev. D} {\bf 66} (2002)
  083005.

\bibitem{CruzOsorio:2010qs} A.~Cruz-Osorio, F.~S.~Guzman and
  F.~D.~Lora-Clavijo, {\it Are scalar field models of dark matter and
    dark energy compatible with a supermassive Schwarzschild black
    hole?}, {\em ArXiv e-prints} (2010)
  [\href{http://xxx.lanl.gov/abs/1008.0027}{{\tt arXiv:1008.0027}}].



\bibitem{Amaro-SeoaneEtAl07} P.~{Amaro-Seoane}, J.~R. {Gair},
  M.~{Freitag}, M.~C. {Miller}, I.~{Mandel}, C.~J. {Cutler}, and
  S.~{Babak}, {\it Intermediate and extreme mass-ratio
    inspirals. Astrophysics, science applications and detection using
    LISA}, {\em Class. and Quant. Grav.} {\bf 24} (2007)
  113.

\bibitem{Barausse:2007}
E.~{Barausse}, L.~{Rezzolla}, D.~{Petroff}, and M.~{Ansorg}, {\it
  {Gravitational waves from extreme mass ratio inspirals in nonpure Kerr
  spacetimes}},  {\em Phys. Rev. D} {\bf 75} (2007) 064026.

\bibitem{Barausse:2008}
E.~{Barausse} and L.~{Rezzolla}, {\it {Influence of the hydrodynamic drag from
  an accretion torus on extreme mass-ratio inspirals}},  {\em Phys. Rev. D}
  {\bf 77} (2008) 104027.

\end{thebibliography}
\end{document}